\documentclass[acmsmall]{acmart}

\usepackage{booktabs} % For formal tables
\usepackage{hyperref}   %gws 20170930
\usepackage{url}  %gws 20170930

\usepackage{multirow}  %gws 20170930
\usepackage{color}  %gws 20170930
\usepackage{diagbox}
\usepackage{amsmath}
\usepackage{arydshln} % gws, 20180202
\usepackage{color} % gws, 20180120

\usepackage{tabularx}
\usepackage{makecell}

\usepackage[linesnumbered,ruled]{algorithm2e} % For algorithms

\SetAlFnt{\small}
\SetAlCapFnt{\small}
\SetAlCapNameFnt{\small}
\SetAlCapHSkip{0pt}
\IncMargin{-\parindent}

\usepackage{soul}
\usepackage{color, xcolor}  %gws 20170930
\sethlcolor{yellow}
\soulregister{\cite}7 % 注册\cite命令
\soulregister{\citep}7 % 注册\citep命令
\soulregister{\citet}7 % 注册\citet命令
\soulregister{\ref}7 % 注册\ref命令
\soulregister{\pageref}7 % 注册\pageref命令
\soulregister{\underline}7 % 注册\underline命令

\acmJournal{jacm}

\setcopyright{acmlicensed}
\begin{document}

% Paper History
% \received{XX 2023}
% \received{XX 2023}

% Title portion
\title{Data Scarcity in Recommendation Systems: A Survey}

\author{Zefeng Chen}
\affiliation{ %College of Cyber Security,
	\institution{Jinan University}
	\city{Guangzhou}
	\country{China}
}
\email{czf1027@gmail.com}

\author{Wensheng Gan}
\authornote{This is the corresponding author.}
\affiliation{
	\institution{Jinan University}
	\city{Guangzhou}
	\country{China}
}
\email{wsgan001@gmail.com}

\author{Jiayang Wu}
\affiliation{ %College of Cyber Security,
	\institution{Jinan University}
	\city{Guangzhou}
	\country{China}
}
\email{csjywu1@gmail.com}

\author{Kaixia Hu}
\affiliation{ %College of Cyber Security,
	\institution{Jinan University}
	\city{Guangzhou}
	\country{China}
}
\email{kaixiahu99@gmail.com}

\author{Hong Lin}
\affiliation{ %College of Cyber Security,
	\institution{Jinan University}
	\city{Guangzhou}
	\country{China}
}
\email{lhed9eh0g@gmail.com}

\begin{abstract}
    The prevalence of online content has led to the widespread adoption of recommendation systems (RSs), which serve diverse purposes such as news, advertisements, and e-commerce recommendations. Despite their significance, data scarcity issues have significantly impaired the effectiveness of existing RS models and hindered their progress. To address this challenge, the concept of knowledge transfer, particularly from external sources like pre-trained language models, emerges as a potential solution to alleviate data scarcity and enhance RS development. However, the practice of knowledge transfer in RSs is intricate. Transferring knowledge between domains introduces data disparities, and the application of knowledge transfer in complex RS scenarios can yield negative consequences if not carefully designed. Therefore, this article contributes to this discourse by addressing the implications of data scarcity on RSs and introducing various strategies, such as data augmentation, self-supervised learning, transfer learning, broad learning, and knowledge graph utilization, to mitigate this challenge. Furthermore, it delves into the challenges and future direction within the RS domain, offering insights that are poised to facilitate the development and implementation of robust RSs, particularly when confronted with data scarcity. We aim to provide valuable guidance and inspiration for researchers and practitioners, ultimately driving advancements in the field of RS.
\end{abstract}


%H.2.8 [Database Applications]: Data mining


\maketitle

\renewcommand{\shortauthors}{Z. Chen \textit{et al.}}

\section{Introduction}

With the rapid development of the Internet \cite{sun2023internet,gan2023web,wan2023web3}, we have entered an era of information explosion \cite{bawden2009dark,gan2017data}. Through the Internet, users can easily access a wide variety of information and content. However, this convenience also brings a serious problem, known as information overload \cite{maes1995agents}. Information overload refers to the overwhelming amount of information available, making it increasingly difficult for individuals to effectively filter and find content that aligns with users' specific interests and preferences. With such an abundance of information, users often find themselves lost in a sea of data, struggling to locate the most relevant and valuable content. It has become increasingly difficult to find content that matches individual interests and preferences in the vast amount of information available. To address this issue, recommendation systems (RSs) \cite{shani2011evaluating, isinkaye2015recommendation, ko2022survey, das2017survey} have emerged as an effective solution. These systems utilize advanced technology and algorithms to analyze user behavior and preferences, providing personalized content or product recommendations to users. By leveraging data on user preferences, browsing history, and purchase behavior, RSs can deliver targeted recommendations that match the individual needs and interests of users. The importance of RSs cannot be overstated. They can greatly improve user experience and satisfaction and have played a crucial role in enhancing business value \cite{schafer2001commerce, fayyaz2020recommendation}. By providing users with personalized recommendations, RSs help to increase user engagement and loyalty, ultimately leading to increased revenue and profitability for businesses. By understanding user preferences and behavior, businesses can deliver targeted recommendations that align with their customers' interests, leading to higher conversion rates, customer satisfaction, and long-term loyalty. In the academic sphere, recommendation systems are also capable of providing recommendations at various levels, encompassing scholarly articles, interpersonal connections, and beyond. These systems leverage sophisticated algorithms and methodologies to offer tailored suggestions that cater to individual preferences and behaviors \cite{deng2020hierarchical}.

In recent times, there has been a notable surge in the research and advancement of RSs \cite{urdaneta2021recommendation, roy2022systematic, maphosa2023fifteen}. Researchers and developers are constantly exploring new approaches and techniques to improve the accuracy and effectiveness of RSs. So far, the RS has been combined with many research fields, including data mining \cite{gan2021survey,fournier2022pattern,gan2021utility}, big data \cite{almohsen2015recommender}, deep learning \cite{zhang2019deep, naumov2019deep}, cloud services \cite{aznoli2017cloud}, and natural language processing (NLP) \cite{shalom2021natural}. The interdisciplinary nature of these advancements has created a wealth of knowledge and expertise in the field, making it an exciting area of study and innovation. As a result, RSs have become indispensable tools in the digital age, addressing the challenges posed by information overload. There is a wealth of knowledge and expertise available in this field, making it an exciting area of study and innovation. It can be said that RSs have revolutionized the way we access and consume information. Through these systems, users can efficiently access information that interests them, saving time and effort. As technology continues to advance, RSs will continue to evolve, providing users with increasingly accurate and tailored recommendations. They could further revolutionize the way we access and consume information in the era of the information explosion. For instance, sequential recommendation, as exemplified by references like \cite{fan2022sequential, fan2023mutual, fan2022sequential2}, represents one of the recommendation methodologies, for predicting a user's next action or item preference in a sequence of interactions. This could involve predicting the next movie a user might watch, the next product they might purchase, or the next article they might read, given their previous behaviors. The underlying assumption is that users' preferences and interests evolve over time, and by considering the sequential patterns in their interactions, more accurate and context-aware recommendations can be generated.

However, despite the significant success of RSs, they still face several major challenges \cite{lu2012recommender}. One of the most significant challenges in this regard is data scarcity, as noted by researchers and practitioners \cite{lu2012recommender, isinkaye2015recommendation}. This issue is particularly pertinent due to the ever-growing volume of available data, which has led to a need for innovative solutions to provide accurate and personalized recommendations. In many applications, the interaction data between users and items is extremely sparse, meaning that we only have limited data to learn about user interests and preferences. Data scarcity poses a significant challenge to the performance of RSs.

Knowledge transfer, a concept central to our discussion, refers to the process of sharing and applying knowledge gained from one context or domain to another. In the context of recommendation systems, it involves leveraging external sources of knowledge to enhance the system's understanding of user behaviors and preferences. In the context of knowledge transfer, addressing the challenge of data scarcity becomes even more crucial. For example, consider a scenario where a new user has just created an account on an e-commerce platform. This user has not yet interacted with any products or items, leading to a complete absence of interaction data. In such cases, traditional recommendation systems may struggle to provide meaningful recommendations due to the lack of user-specific data. Leveraging external sources of knowledge, such as pre-trained language models, could provide a valuable solution to mitigate the impact of data scarcity on RSs. By transferring knowledge from these external sources, RSs can enhance their understanding of user behaviors and preferences, even in situations where direct interaction data is sparse. This approach could potentially bridge the gap when recommending to new users or items with limited interaction history, making the task less daunting \cite{yin2020overcoming}. Furthermore, by incorporating knowledge transfer techniques, RSs have the potential to provide more accurate and meaningful recommendations, ultimately improving their accuracy and personalization \cite{natarajan2020resolving}. This enhanced recommendation quality can better cater to users' needs and preferences, which is not only beneficial for user satisfaction but also plays a pivotal role in driving the growth of Internet enterprises \cite{behera2020personalized}. In essence, techniques including knowledge transfer offer a promising avenue for addressing the challenges posed by data scarcity in RSs, enhancing their recommendation capabilities, and contributing to the overall advancement of the field.

The main contributions of this up-to-date review article are as follows:

\begin{itemize}
    \item We delve into various aspects of RSs, providing detailed insights into their function, classification, framework, and composition. We also highlight the concept of data scarcity and its impact specifically on RSs.

    \item Furthermore, we offer a comprehensive overview of methods and solutions to address the issue of data scarcity in RSs. These include leveraging approaches such as data augmentation, self-supervised learning, transfer learning, broad learning, and knowledge graphs.

    \item Moreover, we explore the challenges and opportunities within the field of RSs, presenting a compelling vision for the future endeavors of researchers and developers.

    \item Finally, we outline the future prospects of RSs and draw some meaningful conclusions.
\end{itemize}

\begin{figure}[h]
    \centering
    \includegraphics[clip,scale=0.42]{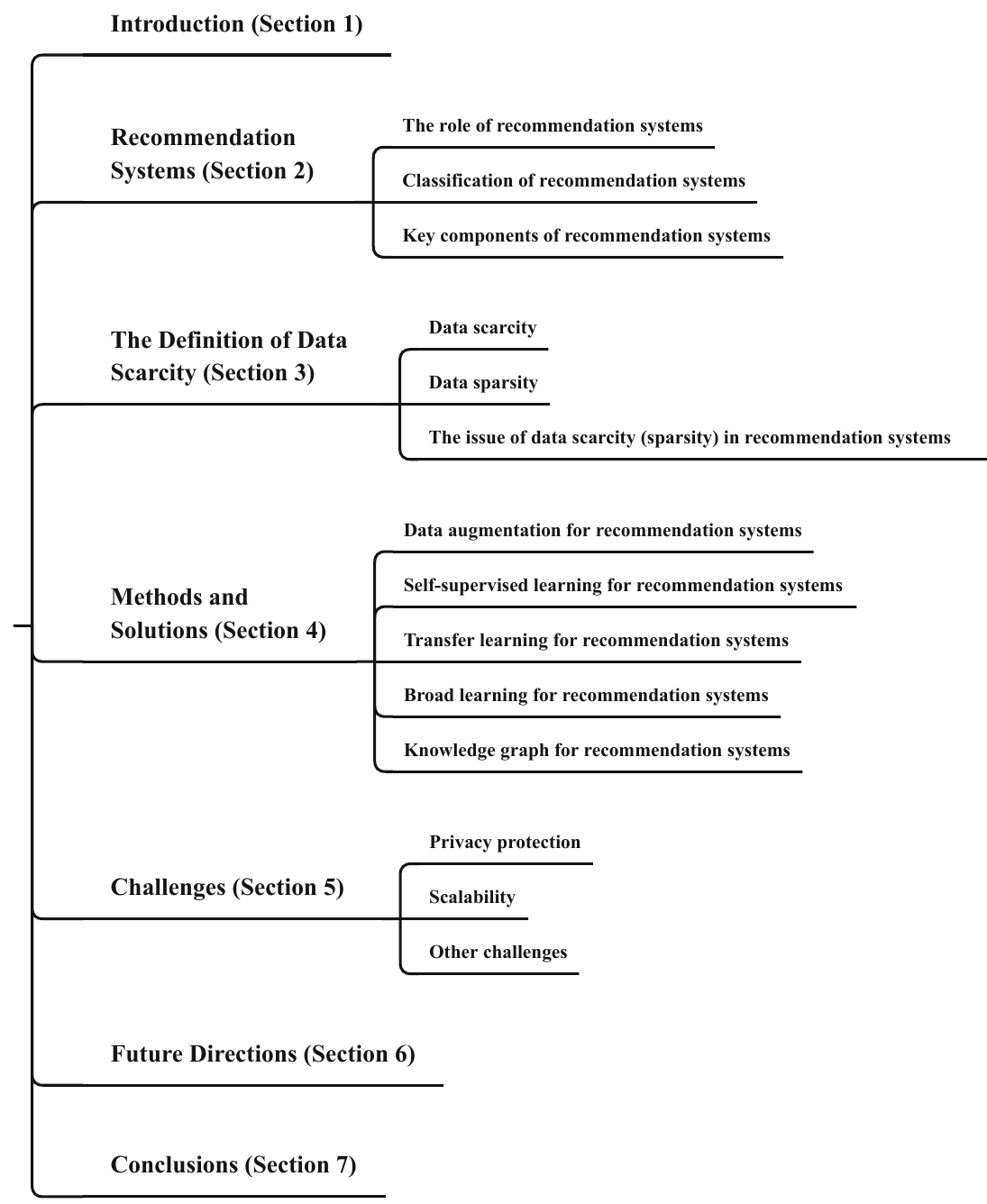}
    \caption{The outline of this review article.}
    \label{fig:outline}
\end{figure}

\textbf{Organization}: The rest of this survey article is organized as follows. In Section \ref{sec:recommendation}, we reviewed and introduced RSs in detail. We present the issue of data scarcity in RSs in Section \ref{sec:scarcity} and list the methods and solutions to this problem in Section \ref{sec:methods}. In Section \ref{sec:challenges} and Section \ref{sec:direction}, we discuss the challenges and future directions in RSs, respectively. Finally, we conclude this paper in Section \ref{sec:conclusions}. The organization of this article is shown in Fig. \ref{fig:outline}.

\section{Recommendation Systems}  \label{sec:recommendation}

Recommendation systems (RSs), also known as recommender systems, belong to the category of information filtering technologies that recommend products, services, or content to users based on their historical behavior, interests, preferences, and contextual information. RSs have gained widespread use in the internet domain, providing personalized recommendations in industries such as e-commerce \cite{schafer1999recommender}, news \cite{karimi2018news}, music \cite{song2012survey}, and movies \cite{phorasim2017movies}, helping users find the content they are interested in amid a vast amount of information \cite{ko2022survey}. The goal of RSs is to enhance the user experience, promote user interaction with content or products, and ultimately achieve business objectives such as increasing user retention and conversion rates.

\subsection{The role of recommendation systems}

The roles and significance of RSs can be examined from both user and company perspectives, as they serve as essential tools for efficiently obtaining information and driving business growth.

i) From the user's perspective, RSs play a crucial role in addressing the challenge of information overload by providing efficient access to relevant and interesting content \cite{wang2021survey}. Due to the abundance of online information, users often feel overwhelmed and struggle to locate their desired content. However, RSs leverage historical data from users to make educated guesses about the content that users may find appealing, particularly when their specific needs are not immediately clear. By analyzing user behavior, preferences, and patterns, these systems can effectively filter and personalize recommendations, helping users discover content that aligns with their interests and preferences. This personalized approach enhances user satisfaction and improves the overall browsing experience by reducing the time and effort required to find relevant information.

ii) From a company's perspective, RSs aim to maximize user attraction, retention, engagement, and conversion rates, thereby facilitating continuous growth in line with the company's business objectives \cite{ricci2015recommender}. Companies with different business models establish specific optimization goals for their RSs. For example, video companies prioritize user watch time, e-commerce companies focus on user purchase conversion rates, and news companies emphasize user click-through rates. It is essential to acknowledge that the core objective of designing a RS is to align with a company's business goals and enhance revenue generation \cite{aggarwal2016recommender}.

Therefore, RSs serve a dual purpose: enabling users to efficiently access the content of interest while simultaneously helping internet companies achieve their business objectives. These dimensions are interconnected and mutually supportive facets of the same issue.

\subsection{Classification of RSs}

The classification of RSs can be distinguished based on recommendation algorithms \cite{sohail2017classifications}. In this way, RSs can be divided into the following four types \cite{narke2020comprehensive} as shown in Fig. \ref{fig:classification}:

\begin{figure}[ht]
    \centering
    \includegraphics[clip,scale=0.34]{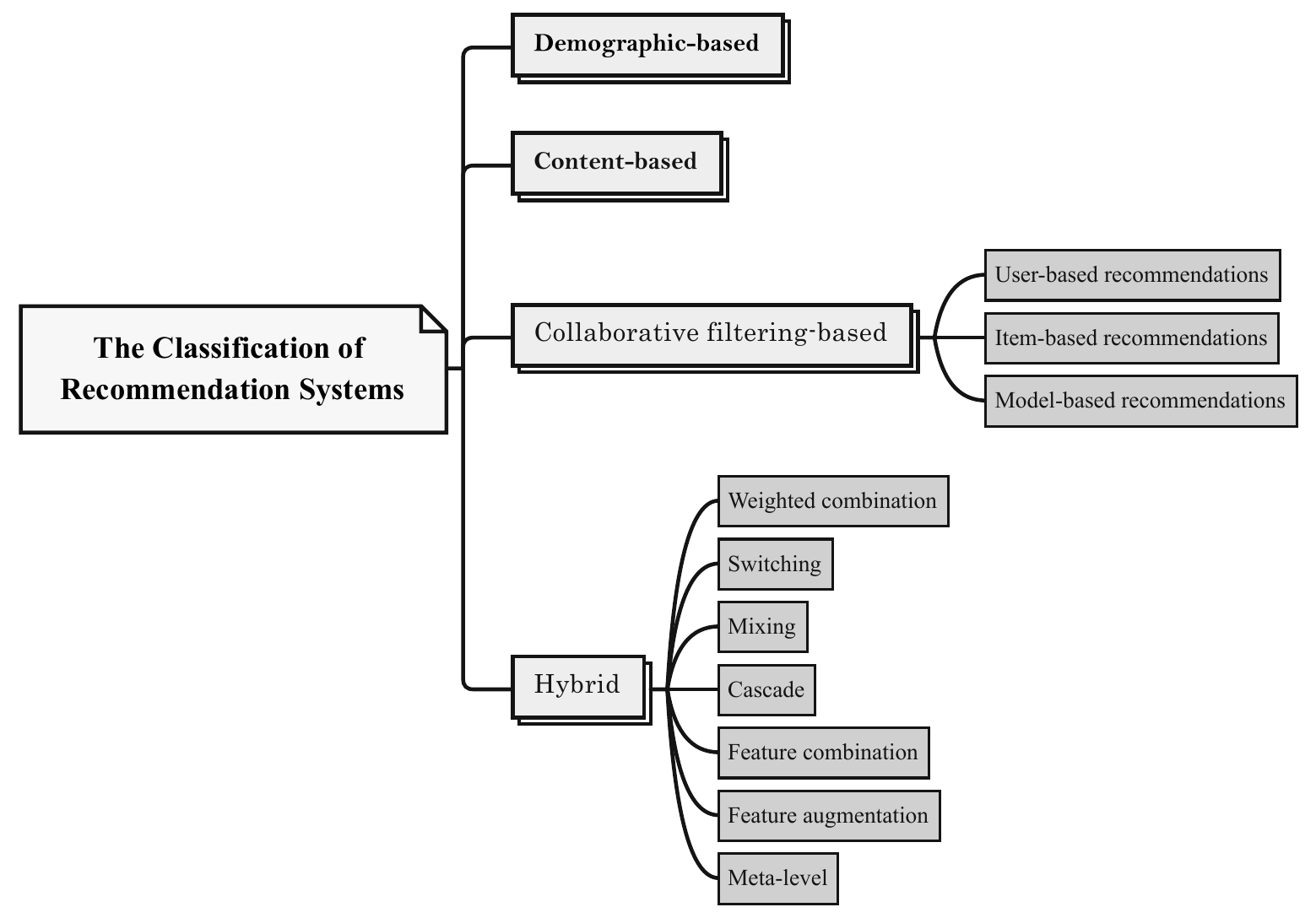}
    \caption{The classification of RSs.}
    \label{fig:classification}
\end{figure}

\textbf{Demographic-based recommendation systems \cite{safoury2013exploiting}.} For demographic-based methods, the user profile records the user's gender, age, active time, etc. as metadata for each user. The basic assumption of this method is that a user is likely to like items liked by similar users. In the process of generating personalized recommendations, RSs compute user similarity by assessing user profiles. Subsequently, they identify the top $K$ users with the highest similarity scores. Afterward, the system utilizes the purchases and ratings of these users to generate recommendations. One popular and straightforward approach is to compile a recommendation list consisting of the items favored by these users. Afterward, the list is arranged according to the average ratings assigned to the items by these selected users. Finally, the system presents the user with an ordered recommendation list.

\textbf{Content-based recommendation systems \cite{pazzani2007content, lops2011content}.}  Content-based methods operate under the assumption that users tend to favor items that share similarities with those they have previously shown interest in. The user's historical records are employed to compute their user profile. The most straightforward recommendation strategy involves calculating the similarity between the user profile and all untried items and then generating a recommendation list in descending order of similarity.

\textbf{Collaborative filtering-based recommendation systems \cite{hameed2012collaborative, suganeshwari2016survey, fan2023graph}.} Collaborative filtering-based recommendations involve gathering users' past behaviors to gain explicit or implicit insights into their preferences for products. This method aims to discover the relevance of items or content itself, or the connections among users, based on their preferences for items or information. Recommendations are then generated using these associations. The user-item preference or rating matrix often consists of a large, sparse dataset. To reduce computational costs, clustering can be applied to either items or users. Content-based recommendations can be broadly categorized into three types: user-based recommendations (identifying users with similar preferences to the target user based on their profiles), item-based recommendations (identifying items that are similar to the ones the target user has previously liked), and model-based recommendations (predicting items that the target user might enjoy based on attributes of the model that considers both users and items).

\textbf{Hybrid recommendation systems \cite{burke2002hybrid, ccano2017hybrid}.} Hybrid RSs are an important research area in recommendation technology. They combine different recommendation techniques to compensate for each other's limitations and achieve better performance. The most common hybrid approach is to combine collaborative filtering with other methods. There are several common hybridization strategies, as shown in Table \ref{table:hybridization strategies}:

\begin{table}[ht]
	\caption{The descriptions of common hybridization strategies.}
	\label{table:hybridization strategies}
	\begin{tabularx}{\textwidth}{m{3.8cm}<{\centering}|m{9.45cm}<{\raggedright}}
		\hline\hline
        \footnotesize\textbf{Hybridization strategies} & \multicolumn{1}{c}{\footnotesize\textbf{Description}} \\
        \hline
        \footnotesize\textbf{Weighted combination} & \footnotesize The predictions from different recommendations are combined through a weighted sum, where the weights are usually optimized. \\
        \hline
        \footnotesize\textbf{Switching} & \footnotesize The switching hybrid strategy involves using different recommendation algorithms in different situations or for different types of users. For example, one algorithm may be more effective for new users, while another algorithm may be better suited for experienced users. The system switches between these algorithms based on user characteristics or contextual factors. \\
        \hline
        \footnotesize\textbf{Mixing} & \footnotesize The recommendation results from different techniques are directly merged to generate the final list. \\
        \hline
        \footnotesize\textbf{Cascade} & \footnotesize Different recommendations are applied sequentially. For example, the content-based recommendation can be used first to solve the cold start problem, then collaborative filtering is applied for refinement. \\
        \hline
        \footnotesize\textbf{Feature combination} & \footnotesize As mentioned earlier, this hybrid strategy involves combining various features and signals to generate recommendations. It considers user preferences, item attributes, social network information, and contextual information to capture a comprehensive understanding of user preferences and provide relevant recommendations. \\
        \hline
        \footnotesize\textbf{Feature augmentation} & \footnotesize The features or profiles from one recommendation are augmented for the input of another. For example, content information can be added to user profiles for collaborative filtering. \\
        \hline
        \footnotesize\textbf{Meta-level} & \footnotesize A meta-model is trained to combine the results from different base recommendations, for example, through ensemble methods.\\
        \hline\hline
        \end{tabularx}
\end{table}

\subsection{Key components of RSs}

Given the elements of user information ($u$), item information ($i$), and scenario details ($s$), the challenge a RS seeks to address can be formally defined as follows: For a user within a particular scenario, the aim is to create a function $f$($u$, $i$, $s$) that anticipates a user's preference for a particular candidate item within a large pool of item information. Subsequently, based on the projected levels of preference, all candidate items are sorted to generate a recommendation list. The definition of the problem leads to an abstract logical framework for RSs. Through the RS, recommendation lists are generated from candidate lists. Although the framework is a broad and formal generalization, it forms the fundamental base upon which the entire technical structure of RSs is developed, with each module being refined and expanded accordingly.

\begin{figure}[ht]
    \centering
    \includegraphics[clip,scale=0.25]{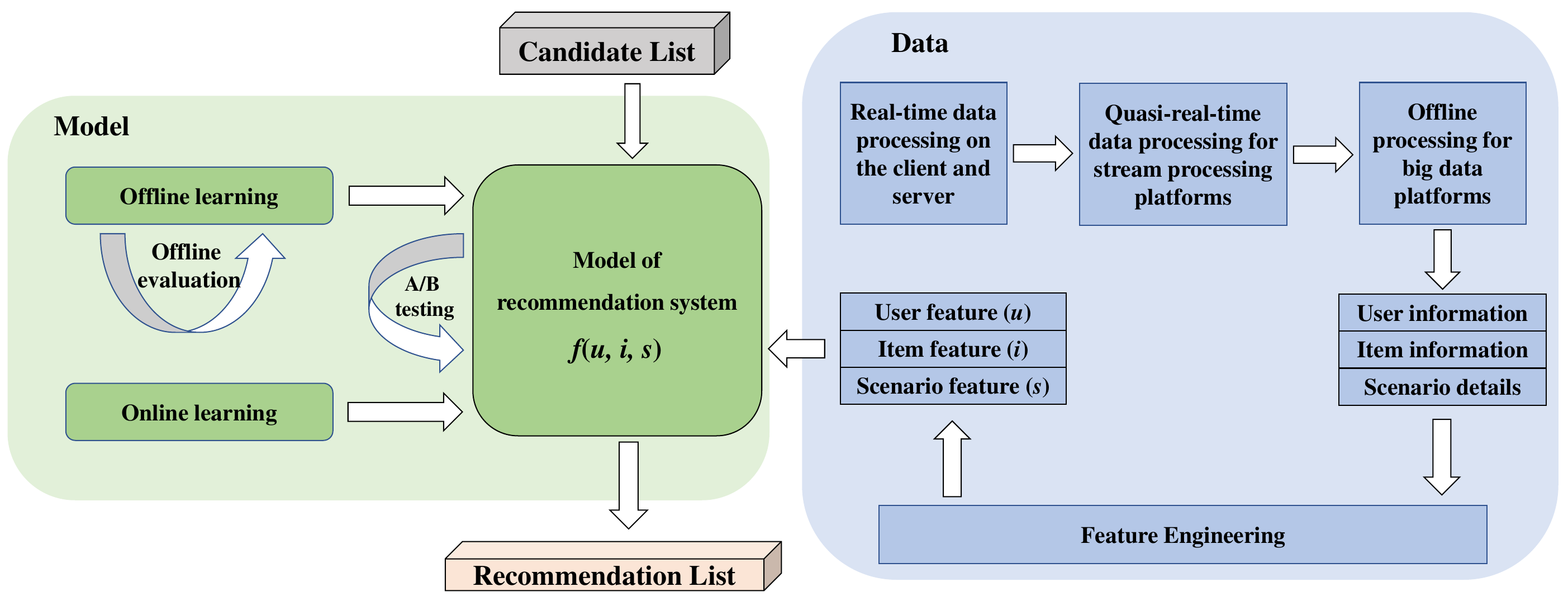}
    \caption{The key components of recommendation systems.}
    \label{fig:components}
\end{figure}

In RSs, there are two main problems to be solved. One is data and information-related issues \cite{peng2022survey}, and another is RS algorithm and model-related issues \cite{naumov2019deep}. The key components of RSs are shown in Fig. \ref{fig:components}. They are used to solve the following problems:

\textbf{Data and information-related issues.} What are the user information, item information, and scenario details? How to store, update, and process them? This part is gradually developing into a data stream framework in RSs that integrates offline batch processing and real-time stream processing of data.

\textbf{Algorithms and model-related issues for RSs.} How to train, predict, and achieve better recommendation results with the recommendation model? This part is refined into a model framework for RSs that integrates training, evaluation, deployment, and online inference.

\subsection{Evaluating indicators of RSs}

To effectively evaluate different RSs, a range of evaluation metrics are required to assess their performance. These metrics help in determining the effectiveness, accuracy, and relevance of the recommendations provided by the system. Here are some key evaluation indicators commonly used in RSs \cite{ko2022survey}, which are shown in Table \ref{table:evaluate}.

\begin{table}[ht]
	\caption{The descriptions of evaluating indicators of recommendation systems.}
	\label{table:evaluate}
	\begin{tabularx}{\textwidth}{m{4.2cm}<{\centering}|m{8.9cm}<{\raggedright}}
		\hline\hline
        \footnotesize\textbf{Evaluating indicators} & \multicolumn{1}{c}{\footnotesize\textbf{Description}} \\
        \hline
        \footnotesize Root Mean Square Error (RMSE) \cite{bag2019noise} & \footnotesize RMSE is a straightforward evaluation metric for RSs, quantifying prediction accuracy through the square root of the average of the squared differences between predicted and actual ratings. It offers insight into the system's precision in forecasting user preferences. \\
        \hline
        \footnotesize Precision \cite{khelloufi2020social} & \footnotesize Precision evaluates the proportion of recommended items that are relevant to the user's preferences, indicating the system's accuracy in suggesting items aligned with user preference. \\
        \hline
        \footnotesize Recall \cite{khelloufi2020social} & \footnotesize Recall calculates the proportion of relevant items that are successfully recommended to the user. It evaluates the ability of the system to cover the user's preferences and ensure that relevant items are not missed. \\
        \hline
        \footnotesize F-measure \cite{liu2015research} & \footnotesize F-measure combines precision and recall using their harmonic mean, offering a balanced evaluation of system performance. It considers both metrics to gauge the overall effectiveness of recommendations. \\
        \hline
        \footnotesize Receiver Operating Characteristic (ROC) Curve \cite{bag2019noise} & \footnotesize The ROC curve demonstrates the trade-off between the false positive rate (FPR) and true positive rate (TPR), aiding comprehension of the precision recall equilibrium. \\
        \hline
        \footnotesize Area Under the Curve (AUC) \cite{krichene2020sampled} & \footnotesize AUC quantifies model performance, with higher values denoting improved recommendation accuracy. \\
        \hline\hline
        \end{tabularx}
\end{table}

\section{The Definition of Data Scarcity}
\label{sec:scarcity}

\subsection{Data scarcity}

Data scarcity refers to the lack of abundant data that can effectively meet the requirements of recommendation systems (RSs) to improve accuracy and prediction dynamics. It entails the challenges in obtaining a sufficient quantity and quality of data in a particular domain or problem, often due to factors like data protection measures and the time-consuming data annotation process. Consequently, acquiring an adequate amount of labeled data becomes difficult, posing a significant obstacle in training accurate RSs \cite{babbar2019data}. It mainly comes from two perspectives:

\textbf{Impact of data protection.} With an increasing focus on privacy and data protection, many organizations and regulations restrict access to and usage of personal and sensitive data. These control measures include enhanced security, privacy safeguards, and licensing requirements to prevent the misuse of personal information. As a result, acquiring a significant amount of diverse data becomes more challenging \cite{leonelli2017data}. Organizations must comply with compliance regulations, which make the data acquisition process more complex and limited, leading to the problem of data scarcity.

\textbf{Time-consuming data annotation.} For many machine learning and artificial intelligence tasks, a considerable amount of labeled data is required for model training and evaluation. However, data annotation is often a time-consuming and resource-intensive task. Manual annotation by skilled professionals is needed, requiring substantial time and effort to carefully analyze the data and assign the correct labels \cite{wang2022whose}. This process may involve numerous data samples, making data annotation time-consuming and costly.

\subsection{Data sparsity}

Data sparsity can lead to scarcity in the distribution of data \cite{yin2020overcoming}. When there is a small amount of data available, the distribution of data in the feature space tends to be uneven, with many uncovered regions or distant data points. This will lead to sparsity in data distribution. Data sparsity can lead to the following issues.

\textbf{Incomplete sample coverage.} Incomplete sample coverage refers to the situation where the available dataset fails to adequately represent the full range of variations and patterns present in the underlying population or phenomenon. This can make it difficult to make predictions or analyze those regions.

\textbf{Imbalanced class distribution.} In classification problems, if some classes have very few samples while others have a larger number of samples, it creates an imbalanced class distribution. This can lead to poor performance when dealing with minority classes since the model has insufficient data to learn from.

\textbf{Challenges in low-density regions.} Data sparsity may lead to regions of low density within the feature space, where there are very few data points. These low-density regions pose challenges for the model, as it may struggle to accurately infer the feature distribution and patterns in those regions from limited data.

\subsection{Issues of data scarcity (sparsity) in RSs}

Data scarcity in recommendation systems refers to the insufficiency of data to effectively capture and model user-item interactions, leading to challenges in providing accurate and personalized recommendations. It often results from factors such as data protection measures and time-consuming data annotation. Data scarcity manifests in several ways, including user sparsity, item sparsity, interaction sparsity, and feature sparsity. Data sparsity, from another aspect, relates to the distribution of data in the feature space, where a small amount of data results in uneven data distribution and incomplete sample coverage. This leads to issues such as imbalanced class distribution and challenges in low-density regions \cite{idrissi2020systematic}. The number of items in RSs often reaches hundreds of thousands or even millions, while user feedback is only available for a small fraction of these items. As a result, the majority of items have very sparse training samples, exhibiting a long-tail property. There are two main reasons for this data sparsity. Firstly, user feedback behavior tends to concentrate on a few popular items, while less popular items receive minimal user feedback. Secondly, RSs typically observe limited user behaviors, meaning that many items lack sufficient user interaction data. Data sparsity can manifest in several aspects:

\textbf{User sparsity.} Some users may only interact with a few items, lacking explicit behavior records for the majority of items. This makes it challenging for the RS to accurately understand the interests and preferences of these users.

\textbf{Item sparsity.} Due to the lack of extensive user behavior data, certain items may only be interacted with by a few users. This poses a challenge for the RS when it comes to personalized recommendations for these items.

\textbf{Interaction sparsity.} The dataset may contain a significant number of missing values, where the interaction records between users and items are incomplete. It can occur due to unrecorded user behavior, data loss, or limitations in data collection.

\textbf{Feature sparsity.} RSs often utilize user and item feature information to assist the recommendation process, but some features may lack data or contain a significant number of missing values. This affects the model's ability to learn and accurately incorporate these features.

Data sparsity poses challenges for RSs, as accurately understanding user interests and item relationships requires reliable and abundant data support. Addressing data sparsity is an important research direction in RSs, requiring the use of appropriate algorithms and strategies to handle missing data and provide accurate and personalized recommendations.

\section{Methods and Solutions} \label{sec:methods}

In the realm of recommendation systems (RSs), the challenge of data sparsity has been a persistent obstacle in providing accurate and personalized recommendations to users. As sparse data limits the system's ability to capture diverse user preferences and item characteristics, researchers and practitioners have been exploring various domain-specific solutions to tackle this issue.

Here, we delve into five distinct and promising approaches that address data sparsity from different angles. Data augmentation for RSs leverages probabilistic density analysis to generate synthetic interactions, augmenting the training data and improving the model's generalization capabilities. Self-supervised learning for RSs enables the model to learn meaningful representations from existing data without explicit labels, uncovering latent user preferences through auxiliary tasks. Transfer Learning for RSs empowers the model to leverage knowledge from related tasks or domains, transferring valuable insights to enhance recommendation performance in sparse data scenarios. Furthermore, broad learning for RSs integrates diverse data sources, enabling a comprehensive understanding of users and items, thus mitigating the impact of data sparsity. Lastly, the knowledge graph for RSs harnesses the power of graph structures and semantic relationships, unearthing hidden patterns and enriching the recommendation process in the presence of sparse data.

\begin{figure}[ht]
    \centering
    \includegraphics[clip,scale=0.33]{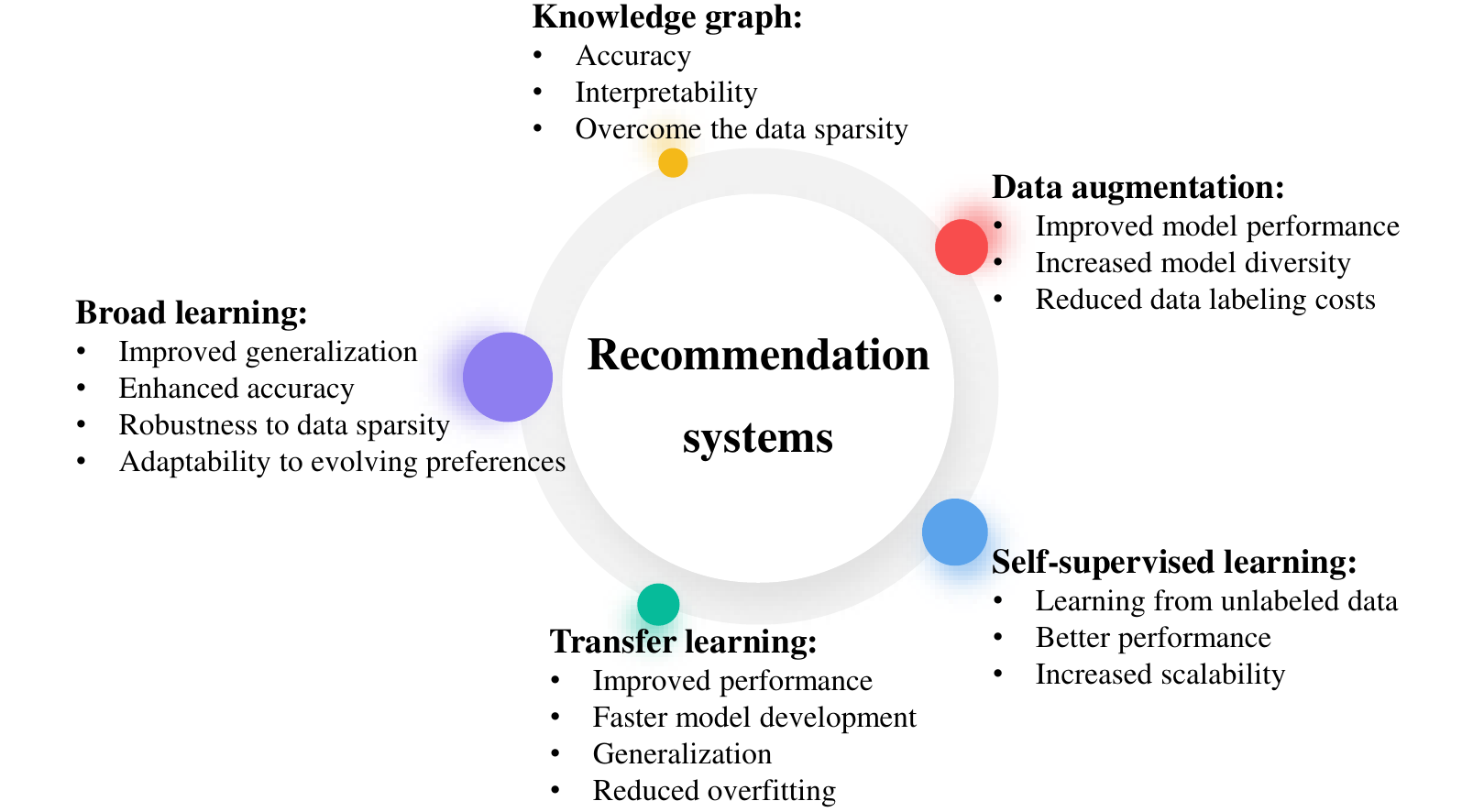}
    \caption{Methods and solutions for recommendation system.}
    \label{fig:method}
\end{figure}

By exploring these domain-specific solutions, we aim to shed light on state-of-the-art techniques and insights that can effectively address data sparsity challenges in RSs, which are shown in Fig. \ref{fig:method}. By embracing these approaches, we can move towards building more robust and accurate RSs, providing users with personalized and meaningful recommendations even in the face of data sparsity.

\textbf{Data augmentation for recommendation systems \cite{bansal2022systematic}.} Data augmentation techniques in RSs aim to alleviate data sparsity by generating additional synthetic interactions based on the existing sparse data \cite{yang2022image}. These methods typically leverage probabilistic density analysis to create new user-item interactions, thereby increasing the volume of training data available for recommendation models. By introducing synthetic interactions, the model can better generalize to unseen scenarios and improve its ability to make accurate and diverse recommendations. For instance, consider a scenario where user-item interactions are sparse, particularly for new items. Data augmentation techniques could generate simulated interactions based on existing data patterns, effectively filling gaps and improving the recommendation accuracy.

\textbf{Self-supervised learning for recommendation systems \cite{yu2023self}.} Self-supervised learning approaches for RSs focus on learning meaningful representations from the existing data without relying on explicit labels \cite{yu2023self}. These methods design auxiliary tasks that encourage the model to capture underlying user preferences and item characteristics. By reconstructing missing interactions, predicting masked elements, or formulating pretext tasks, self-supervised learning enhances the model's ability to understand and leverage the available data for recommendation. For example, imagine a situation where a user's explicit preferences are scarce, but their implicit behavior (e.g., the types of items they click on or dwell upon) is more abundant. Self-supervised learning could help uncover latent user preferences through auxiliary tasks, improving recommendation accuracy.

\textbf{Transfer learning for recommendation systems \cite{zhao2013active}.} Transfer learning techniques in RSs aim to improve recommendation performance in situations where data is sparse. This is achieved by leveraging knowledge from related tasks or domains \cite{pan2016survey}. The process involves initially training the model on a source domain or task, and subsequently fine-tuning it on the target domain using the limited data available. By employing this approach, the model can adapt and generalize more effectively to the new scenario, resulting in more accurate and effective recommendations through the transfer of knowledge from the source to the target domain. For instance, consider a scenario where a new e-commerce platform lacks sufficient user-item interaction data. By transferring knowledge from a similar, more established e-commerce platform, the RS can enhance its recommendations by utilizing insights from the well-established platform.

\textbf{Broad Learning for recommendation systems \cite{zhu2017broad}.} Broad learning methods tackle data sparsity in RSs by integrating diverse data sources, such as user behavior logs, user profiles, and item features \cite{zhu2017broad}. These approaches seek to obtain a comprehensive understanding of users and items by considering multiple data modalities. By combining different data sources, broad learning enhances the ability of the model to capture complex user preferences and item characteristics, leading to improved recommendation accuracy in sparse data scenarios. For example, in a scenario where user interactions are limited, broad learning could incorporate additional information such as user demographics, item attributes, or external data sources to mitigate the impact of data sparsity.

\textbf{Knowledge graph for recommendation systems \cite{shao2021survey}.} Knowledge graph-based methods address data sparsity by leveraging the explicit semantic relationships and additional information captured in the knowledge graph \cite{guo2020survey}. These methods utilize graph-based models or knowledge graph embeddings to effectively exploit the structural information in the graph. By mining and integrating knowledge from the graph, RSs can discover hidden patterns and relationships that are challenging for traditional methods, thus enhancing the quality and diversity of recommendations. For instance, consider a movie recommendation system where user-item interactions are sparse. A knowledge graph could capture relationships between movies, actors, genres, and user preferences, enriching the recommendation process by recommending movies based on related attributes and user interests.

There are detailed introductions to five techniques and how they empower RSs listed as follows. These methods play crucial roles in empowering RSs, enabling them to provide personalized and relevant recommendations to users.

\subsection{Data augmentation for RSs}
\subsubsection{Definition}

Data augmentation is a data space solution that can be utilized to tackle the problem of data scarcity in RSs \cite{bansal2022systematic}. It is a frequently employed technique in machine learning and deep learning to augment training datasets by creating new instances or modified versions of the initial data \cite{van2001art, chlap2021review}. The objective of data augmentation is to enhance the performance and generalization ability of a model of machine learning by exposing it to a broader range of variations and examples derived from the existing data.

Data augmentation for RSs \cite{song2022data} refers to the process of generating additional training data by applying transformations or manipulations to the existing data to enhance both the diversity and quantity of the available training examples. It is done to improve the accuracy and robustness of recommendation models, particularly in cases where the original dataset is limited, biased, or unrepresentative of the target population.

\subsubsection{Advantages}

As one of the solutions to the scarcity of RSs, data augmentation for RSs offers several advantages:

\begin{itemize}
    \item  \textbf{Improved model performance \cite{chlap2021review}.} By generating additional training data, data augmentation can enhance the precision and resilience of recommendation models, particularly in cases where the original dataset is limited, biased, or unrepresentative of the target population.

    \item  \textbf{Increased model diversity \cite{gao2021enabling}.} By increasing the diversity of the training examples, data augmentation can help reduce overfitting and improve the generalization capability of recommendation models.

    \item \textbf{Reduced data labeling costs \cite{lei2019preliminary}.} Data augmentation can help reduce the need for manual data labeling and annotation, which can be time-consuming and costly.

    \item \textbf{Better handling of class imbalance \cite{temraz2022solving}.} Data augmentation can be used to address class imbalance issues in recommendation datasets, where certain items or users may be significantly overrepresented or underrepresented.
\end{itemize}

\subsubsection{Methods}

\textbf{Dataset augmentation.} One of the techniques of data augmentation is dataset augmentation, DeVries and Taylor \cite{devries2017dataset} proposed a simpler, domain-agnostic approach to dataset augmentation in feature space. The technology used in supervised learning, and can be applied to RSs to increase the variability of datasets and ultimately produce more robust RS models.

\textbf{Smart augmentation learning.} Lemley \textit{et al.} \cite{lemley2017smart} proposed a strategy for optimal data augmentation, which is smart augmentation learning. Smart augmentation learning involves training a network to generate augmented data during the training process of the target network. The generated data is designed to minimize network loss, thereby enhancing the performance of the target network. Furthermore, this strategy has demonstrated the potential to attain comparable or superior levels of performance with significantly reduced network sizes across a range of tested scenarios. It provides ideas for augmenting data in the network within the RS model.

\textbf{Auto augment.} In work \cite{cubuk2018autoaugment}, the simple procedure called AutoAugment is described as being able to automatically seek enhanced data augmentation strategies. In order to increase the speed of auto augment, Lim \textit{et al.} \cite{lim2019fast} proposed a more efficient search strategy based on density matching to find effective augmentation policies, and an improved algorithm was introduced, resulting in several orders of magnitude reduction in search time. The procedure of auto augment is suitable for RSs to enable automatic data augmentation.

\textbf{Population-based augmentation.} AutoAugment is computationally infeasible for ordinary users. Ho \textit{et al.} \cite{ho2019population} propose a novel data augmentation technique called Population-Based Augmentation. Unlike traditional approaches that utilize fixed augmentation policies, this method generates non-stationary augmentation policy schedules. By introducing dynamic policies, a slight reduction in error is observed, indicating improved performance. It provides feasible data augmentation solutions for ordinary users in RSs.

\textbf{Adversarial AutoAugment \cite{zhang2020adversarial}.} It is a technique that can optimize both the objective-related object and the augmentation search loss at the same time. By generating adversarial augmentation policies, the expanded policy network seeks to elevate the training loss of the target network, enabling it to acquire more resilient features from demanding instances and enhance its overall generalization. Similarly, this also provides a reference strategy for RSs to achieve adversarial augmentation of data.

Recently, a novel work of data augment for the RS has emerged \cite{song2022data}. In this work, a set of data augmentation strategies for the sequential recommendation was proposed. The technique can enhance the model's ability to generalize, particularly in cases where the training data is limited, resulting in a substantial improvement in the model's performance. It can be expected that more and more data augmentation techniques will be applied to RSs to effectively address the data scarcity problem.

\subsection{Self-supervised learning for RSs}
\subsubsection{Definition}

Self-supervised learning is a machine learning technique where a model is trained on unlabeled data to acquire meaningful data representations \cite{liu2021self}. Unlike supervised learning, which relies on labeled data with correct output or target values, labeled data for training is not required in self-supervised learning \cite{zhai2019s4l}. Instead, it creates a supervised task using the input data itself, where the model is trained to make predictions about one portion of the input data using another portion \cite{liu2021self}. For instance, in image self-supervised learning, the model is trained to predict the absent part of an image when provided with the remaining portion as input. Similarly, in NLP, the model may predict the next word in a sentence based on the preceding words. By training on such self-supervised tasks, the model learns to capture significant characteristics or representations from the input data, which can then be utilized for various downstream tasks like classification or regression. These learned representations serve as compressed versions of the input data, capturing the most crucial information relevant to the given task.

Self-supervised learning can also be a useful technique for addressing data scarcity in RSs \cite{huang2022self, yu2023self}. RSs often struggle to perform well when there is a scarcity of labeled data available. This is because obtaining labeled data for RSs can be challenging, as it requires collecting feedback from users on their past interactions with the system, which can be time-consuming and expensive. Self-supervised learning can be a useful technique for addressing this challenge, as it allows models to learn useful representations of the available unlabeled data, such as user-item interactions or content features, which can then be used to make recommendations.

\subsubsection{Advantages}

To deal with the issues of data scarcity, self-supervised learning in RSs has several advantages as follows:

\begin{itemize}
    \item \textbf{Learning from unlabeled data \cite{liu2019exploiting}.} Self-supervised learning, leveraging unlabeled data, reduces annotation costs while enhancing generalization. It excels in resource scarcity, multitasking learning, and data scarcity scenarios. By learning informative representations from data, self-supervised learning equips models to adapt to multiple tasks, potentially advancing the field of RS.

    \item \textbf{Better performance \cite{hendrycks2019using}.} By employing self-supervised learning on a vast amount of unlabeled data, a model can acquire more pertinent and valuable features or representations of the data. These can subsequently be fine-tuned on a smaller labeled dataset for recommendation. This approach can yield improved performance on the recommendation task, even when faced with limited labeled data.

    \item \textbf{Increased scalability \cite{wu2020tracklet}.} Self-supervised learning offers the opportunity to pre-train models using vast amounts of unlabeled data, which can subsequently be fine-tuned on smaller labeled datasets to address specific recommendation tasks. This approach can lead to increased scalability, as it allows models to generalize well to new and unseen data.
\end{itemize}

\subsubsection{Methods}

\begin{figure}[ht]
    \centering
    \includegraphics[clip,scale=0.33]{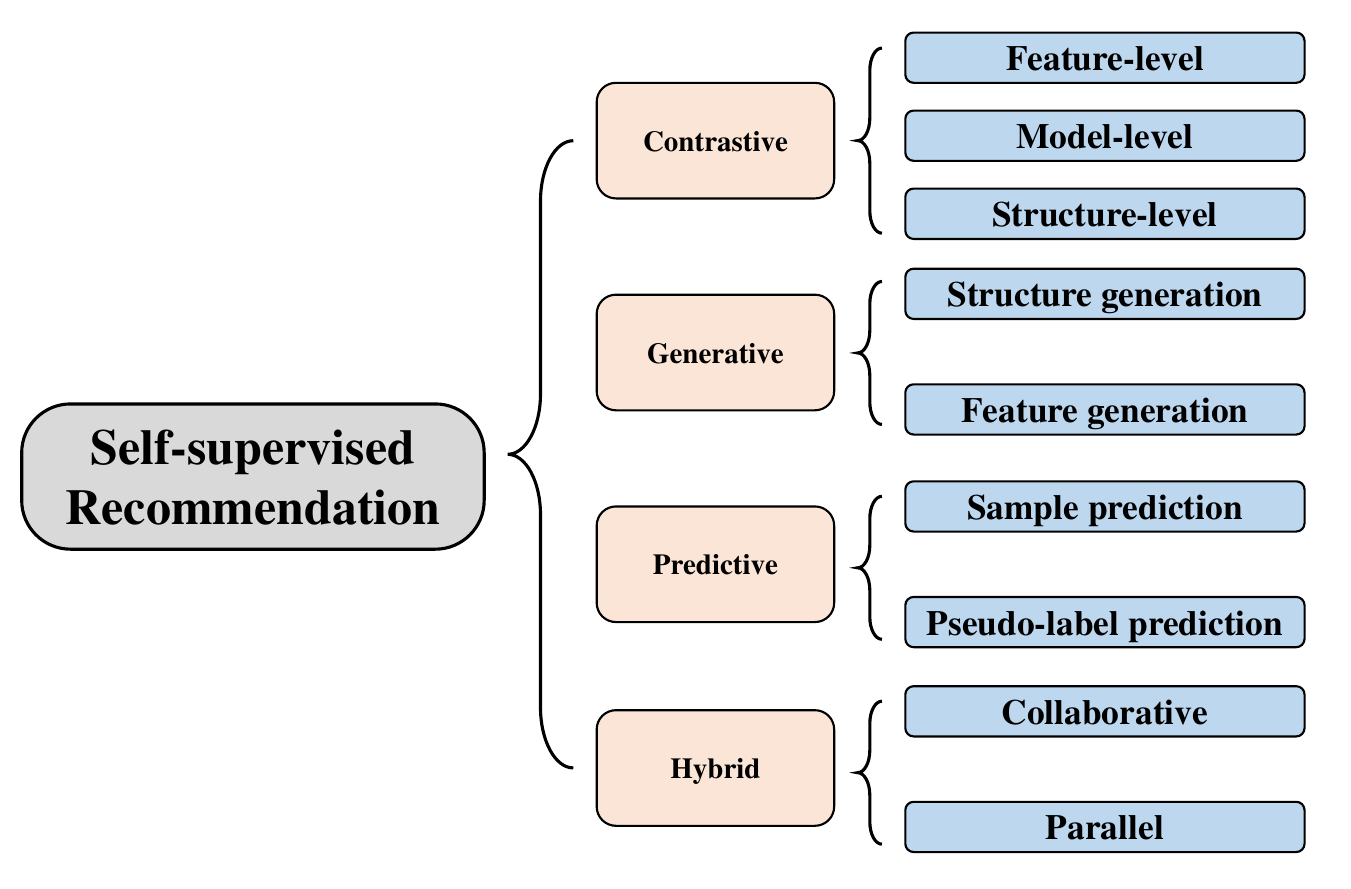}
    \caption{The taxonomy of self-supervised recommendation.}
    \label{fig:self}
\end{figure}

Self-supervised learning for recommendation models can be classified into four categories based on the nature of their self-supervised tasks: contrastive, predictive, generative, and hybrid. This distinguishes self-supervised learning for RSs from other paradigms \cite{yu2023self}. The taxonomy of self-supervised recommendation is illustrated in Fig. \ref{fig:self}.

\textbf{Contrastive methods.} Contrastive methods \cite{jaiswal2020survey}, which are a prominent aspect of self-supervised learning, leverage contrastive learning as their driving force. These methods treat each instance, such as users, items, and sequences, as distinct classes. The core concept is to reduce the distance between different versions of the same instance in the embedding space and increase the separation between versions of different instances. To generate these variants, various transformations are applied to the original data. These variants must introduce non-essential variations without significantly altering the original instance. Positive pairs consist of two variants of the same instance, while negative pairs consist of variants of different instances. Contrastive methods aim to obtain discriminative representations for recommendation tasks by enhancing consistency within positive pairs and reducing the concurrence among negative pairs.

\textbf{Generative methods.} Generative techniques \cite{liu2021self} within self-supervised learning for recommendation focus on creating fresh samples from existing data. These strategies are influenced by masked language models \cite{salazar2020masked} like BERT \cite{devlin2019bert}, proven effective in various natural language processing (NLP) endeavors. In the recommendation context, user/item profiles are initially distorted, and the generative model is then trained to reconstruct these profiles from their distorted versions. This self-supervised task helps the model learn meaningful representations of the data, which can then be used for recommendation. The most common tasks in generative approaches are structure and feature reconstruction. In structure reconstruction, the model is trained to predict absent values in the input data, such as missing ratings or user/item attributes. Feature reconstruction, for another prospect, involves generating new features that are relevant to the recommendation task. For example, a generative model can learn to generate new user/item profiles by combining information from multiple existing profiles.

\textbf{Predictive methods.} In self-supervised representation learning \cite{hsu2021hubert}, predictive and generative methods fulfill distinct roles. Generative methods aim to predict missing portions of the initial data, whereas predictive methods generate new samples or labels from the existing data to aid the pretext task. Predictive techniques fall into two categories: sample-based and pseudolabel-based methods. Sample-based methods predict informative samples using the current encoder parameters and employ these predictions to create new, more confident samples. This approach combines self-training and self-supervised learning, both forms of semi-supervised learning. Pseudolabel-based methods utilize a generator, which can be either another encoder or rule-based selectors, to create labels. These generated labels are subsequently employed as reference points for guiding the encoder.

\textbf{Hybrid methods.} To achieve comprehensive self-supervision in recommendation models, it is advantageous to combine different types of self-supervised learning methods that leverage diverse self-supervision signals. Hybrid methods frequently employ multiple encoders and projection heads, enabling multiple self-supervised tasks to work simultaneously or collaborate to enhance self-supervision signals. Merging diverse pretext tasks often entails assigning weights to distinct self-supervised losses across categories, diminishing redundancy, and bolstering the recommendation model's global performance.

\subsection{Transfer learning for RSs}
\subsubsection{Definition}

Transfer learning is a valuable machine learning technique that enables a model trained on one task to be applied to a related but different task \cite{ribani2019survey}. Instead of training a model from scratch on a new task with a large amount of data, transfer learning takes advantage of the knowledge and learned representations from a pre-trained model \cite{marcelino2018transfer}. The pre-trained model, typically trained on extensive datasets like ImageNet for image recognition or Word2Vec for natural language processing, has already acquired valuable features and patterns. These learned features can be transferred to a new task by either using the pre-trained model as a fixed feature extractor or fine-tuning some or all of its layers on the new task. Transfer learning is particularly beneficial when the new task has limited available data or when training a model from scratch is computationally demanding. By leveraging the pre-trained model, transfer learning can enhance performance and accelerate the training process on the new task.

\begin{figure}[ht]
    \centering
    \includegraphics[clip,scale=0.2]{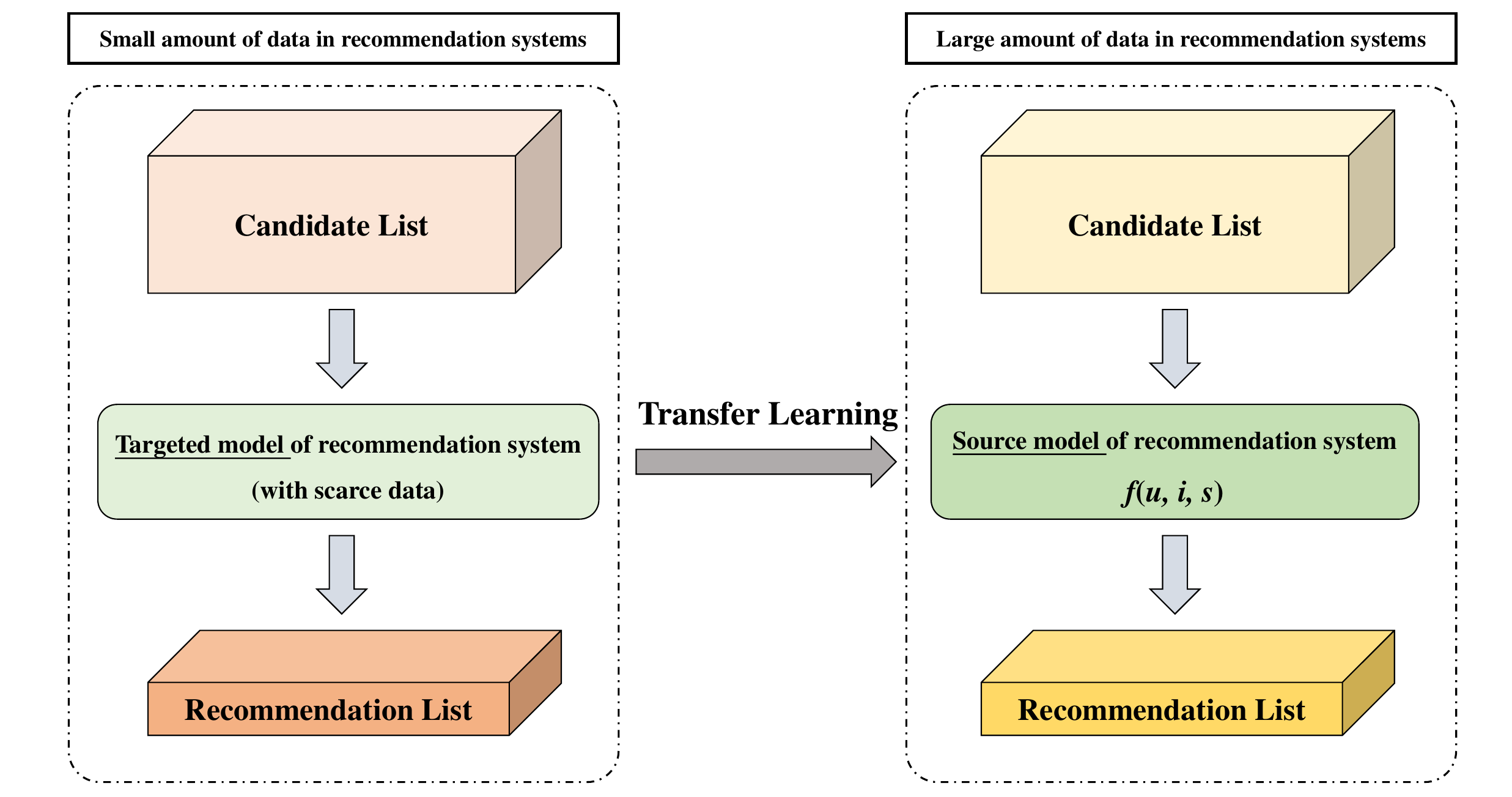}
    \caption{The schematic diagram of transfer learning in RSs.}
    \label{fig:transfer}
\end{figure}

\subsubsection{Advantages}

In the context of RSs, transfer learning can offer several advantages when data scarcity is an issue. Here are some of those advantages:

\begin{itemize}
    \item \textbf{Improved performance \cite{moreno2012talmud, wang2019recsys}.} By leveraging knowledge learned from a source domain with abundant data, transfer learning can enhance the performance of an RS in a target domain with limited data. It is particularly beneficial when there is not enough data available to train a reliable recommendation model from scratch.

    \item \textbf{Faster model development \cite{ma2012transfer}.} Transfer learning allows the pre-trained model from the source domain to be used as a starting point for the target domain. This approach saves time and resources by minimizing the requirement for extensive training on limited data. Developers can focus on fine-tuning the model to adapt it to the target domain, instead of building an entirely new model from scratch.

    \item \textbf{Generalization \cite{zhuang2020comprehensive}.} Transfer learning helps in generalizing recommendations across domains. It allows the model to capture common patterns, preferences, and relevant relationships in the source and target domains. It enables the model to make accurate recommendations even with limited data in the target domain.

    \item \textbf{Reduced overfitting.} Limited data can often lead to overfitting, where a model becomes too specialized to the training data and fails to generalize well to new instances. Transfer learning helps alleviate overfitting by leveraging the knowledge gained from the source domain. The pre-trained model provides a regularization effect, as it has learned from diverse data and can prevent the model from relying too heavily on the limited target domain data.
\end{itemize}

\subsubsection{Methods}

Cross-system recommendation and cross-domain recommendation are two kinds of recommendation methods. Cross-system recommendation refers to recommending items between different RSs. Different RSs may have different data sources, algorithms, and objective functions. Therefore, when recommending items between different systems, data transformation, and adaptability need to be considered. Cross-system recommendation can help users obtain personalized services on different platforms and applications. Cross-domain recommendation refers to applying users' interests and behavior information in one domain to recommend items in another domain. For example, applying users' interests and behavior information in the shopping domain to recommend movies in the movie domain. The cross-domain recommendation could help users discover new interest areas and expand their consumption and experience scope.

We introduce the methods of transfer learning in cross-system recommendation and cross-domain recommendation.

\textbf{Cross-system recommendation transfer learning.} Cross-system recommendation transfer learning refers to the process of transferring knowledge from one RS to another to enhance the recommendation performance of the target system. It involves leveraging insights gained from one domain, such as movies or e-commerce, and applying them to another domain, such as music or social networks. Shared feature or model fusion techniques are commonly employed in cross-system recommendation transfer learning to facilitate knowledge transfer. For instance, Zhao \textit{et al.} \cite{zhao2013active} proposed an active learning-based framework that establishes entity correspondences between RSs with limited resources, enabling effective knowledge transfer. They introduced a unified framework for proactive cross-system recommendation transfer learning. This framework employs active learning principles to establish entity correspondences across different systems, enabling flexible and cost-effective knowledge transfer \cite{zhao2017unified}.

\textbf{Cross-domain recommendation transfer learning.} Cross-domain recommendation transfer learning involves leveraging insights from one domain to improve recommendation performance in another domain. For example, it involves leveraging knowledge from movie recommendations to improve book recommendations or using insights from music recommendations to enhance restaurant recommendations. In cross-domain recommendation transfer learning, methods including pre-trained models or shared feature methods are commonly employed to facilitate knowledge transfer. Several approaches have been developed in this area. For instance, TALMUD \cite{moreno2012talmud} is a transfer learning approach that addresses multiple domains. Lu \textit{et al.} \cite{lu2013selective} proposed an improved framework that integrates an empirical prediction error and variance criterion for selective knowledge transfer in cross-domain recommendation. Chen \textit{et al.} \cite{chen2017tlrec} proposed TLRec, which utilizes overlapping users and items as bridges to facilitate knowledge transfer between different domains in cross-domain RSs. Krishnan \textit{et al.} \cite{krishnan2020transfer} introduced a technique using contextual invariants for the one-to-many cross-domain recommendation. Li \textit{et al.} \cite{li2020atlrec} presented an attentional adversarial transfer learning network to enhance cross-domain collaborations and interdisciplinary knowledge exchange. Additionally, Li \textit{et al.} \cite{li2020ddtcdr} proposed a novel cross-domain recommendation approach employing dual learning, iteratively transferring information between interconnected domains. These approaches demonstrate the potential of cross-domain recommendation transfer learning to improve recommendation performance by leveraging knowledge from related domains.

\subsection{Broad learning for RSs}
\subsubsection{Definition}

Broad learning, also known as breadth-first learning, is a machine learning approach that focuses on exploring a wide range of features or input dimensions \cite{chen2017broad, zhang2019broad, gong2021research}. By considering diverse information from various sources, such as different types of data, broad learning aims to improve learning performance and enhance the model's ability to generalize. It offers a powerful way to handle high-dimensional and heterogeneous data, capturing different patterns and characteristics for a more comprehensive understanding \cite{cai2018feature}.

Hence, applying broad learning to RSs can improve their effectiveness and accuracy. By considering a wide range of features and dimensions, broad learning captures comprehensive user preferences and item characteristics. It expands beyond traditional approaches by incorporating diverse information such as demographics, social connections, and contextual data. Broad learning enhances recommendation accuracy and adapts to changing preferences, providing more personalized and relevant recommendations for users. By leveraging a wider set of features and dimensions, broad learning enhances the system's ability to provide personalized and accurate recommendations, making it possible to be a valuable approach in the field of RSs.

\subsubsection{Advantages}

When it comes to RSs facing data scarcity, broad learning provides several significant advantages. These advantages stem from the ability of broad learning to leverage a diverse range of features and dimensions, enabling the system to overcome the limitations imposed by limited data availability.

\begin{itemize}
    \item \textbf{Improved generalization \cite{huang2019motor}.} In data-scarce scenarios, traditional recommendation models may struggle to generalize well due to limited training data. Broad learning, by considering a wide range of features and dimensions, helps overcome this limitation. It incorporates diverse information and captures a more comprehensive understanding of user preferences and item characteristics, leading to improved generalization and better recommendation performance.

    \item \textbf{Enhanced accuracy \cite{huang2022accurate}.} By incorporating a broader set of features and dimensions, broad learning can capture more nuanced user preferences and item characteristics. This enables the RS to make more accurate predictions and provide personalized recommendations even with limited data. The inclusion of additional contextual information can help compensate for the lack of explicit user feedback and improve recommendation accuracy.

    \item \textbf{Robustness to data sparsity.} Data scarcity often leads to sparse user-item interaction matrices in RSs. Broad learning can mitigate the impact of data sparsity by leveraging a wider range of features and dimensions. It can capture various aspects of user behavior, such as demographics, social connections, and contextual information, which can help fill the gaps left by sparse user-item interactions. It leads to more robust and reliable recommendations, even when the available data is limited.

    \item \textbf{Adaptability to evolving preferences \cite{zhou2019preference}.} In data-scarce scenarios, user preferences may change over time, making it challenging to capture and adapt to these changes. Broad learning, by considering a diverse set of features and dimensions, can capture evolving user preferences more effectively. It allows the RS to adapt and update recommendations based on new information and changing user behavior, ensuring that the recommendations remain relevant and up-to-date.
\end{itemize}

\subsubsection{Methods}
The existing broad learning methods for RSs can be mainly divided into feature-based, model-based, data-based, and task-based broad learning.

\textbf{Feature broad learning.} It aims to enhance the model's expressive power by introducing more features. Traditional machine learning models typically use a limited number of features for modeling. Feature broad learning can capture more diverse and comprehensive user preferences and item features by introducing additional feature dimensions such as user behavior, contextual information, and social relationships. It enhances the model's accuracy and generalization performance. By introducing more feature dimensions, feature broad learning helps enrich the model's description of users and items, thereby alleviating data sparsity issues. For example, when a user has limited interaction data, additional features such as user profiles and interest tags can be considered to provide a more comprehensive user representation. In 2022, a novel multi-label classifier \cite{huang2022accurate} was applied in broad learning to improve accuracy and training efficiency, which is well-suited for large-scale multi-label learning within feature broad learning scenarios.

\textbf{Model broad learning.} It focuses on improving prediction performance by combining multiple models or learners. It leverages different model structures, algorithms, or learning strategies and integrates them into a unified model to achieve stronger recommendation capabilities. By integrating multiple models or learners, model broad learning leverages their strengths to compensate for data sparsity issues. Different models can capture data from different perspectives, and by combining their predictions, recommendation accuracy and robustness can be improved. For example, the model of the broad learning system \cite{chen2017broad} provides an alternative approach for deep learning and structure, and the regularized robust broad learning system \cite{jin2018regularized} is developed to enhance the robustness of the system. In addition, the fuzzy broad system \cite{feng2018fuzzy} and the stacked broad learning system \cite{liu2020stacked} can achieve higher performance with fewer rules and less runtime.

\textbf{Data broad learning.} It aims to enhance the model's learning ability by incorporating additional data sources or types. Conventional RSs typically depend on user-item interaction data for their modeling, while data broad learning combines other types of data (e.g., user social data, text data, and image data) with interaction data to provide more comprehensive and diverse information, thereby improving recommendation accuracy and personalization. By incorporating additional data sources or types, data broad learning enhances the model's learning ability to better handle data sparsity. For example, combining user social data or text data can provide more information about user interests and preferences, leading to improved recommendation effectiveness.

\textbf{Task broad learning.} Task broad learning aims to improve model performance by simultaneously learning multiple related tasks. Traditional RSs typically address a single recommendation task, such as item recommendation or rating prediction. Task broad learning enables the simultaneous learning of multiple relevant recommendation tasks, such as item recommendation and user interest prediction. Leveraging shared, and transfer learning enhances the model's generalization ability and recommendation effectiveness. By simultaneously learning multiple related tasks, task broad learning utilizes the correlation between tasks to enhance model performance. When one task has high data sparsity, the information from other tasks can provide assistance and supplementation, improving the model's performance on sparse tasks. Zhu \textit{et al.} \cite{zhu2017broad} proposed a multi-source collaborative recommendation based on task-based broad learning. The collaboration of multi-source recommendation tasks enables breadth learning to make better recommendations based on tasks.

\subsection{Knowledge graph for RSs}
\subsubsection{Definition}

A knowledge graph is an information system that organizes and represents knowledge using the structure of a graph. It models entities, concepts, and relationships from the real world and displays their associations in the form of a graph. Knowledge graphs typically consist of nodes (representing entities or concepts) and edges (representing the relationships between them). A knowledge graph-based RS leverages the information from a knowledge graph to provide personalized recommendations. Unlike traditional RSs based on user behavior or item attributes, knowledge graph-based RSs utilize domain knowledge, entity relationships, and semantic information from the knowledge graph to better understand user interests and item features.

\subsubsection{Advantages}
For a knowledge graph-based RS, integrating the technique of knowledge graphs as auxiliary information brings three main advantages.

\textbf{Accuracy.} It improves the accuracy of the RS. By leveraging the entity relationships and semantic information from the knowledge graph, the RS gains a more comprehensive understanding of user interests and item features. Thus, it leads to more precise and personalized recommendations. For example, the system can use the relationships in the knowledge graph to discover related items, even if these items have not been explicitly expressed in the user's historical behavior.

\textbf{Interpretability.} A knowledge graph-based RS offers interpretability. The knowledge graph visualizes the relationships between entities and provides explanations for the recommended results. This allows users to understand why certain items are recommended to them and the associations between those items and their interests and preferences. Interpretability helps build trust in the RS and provides a better user experience.

\textbf{Overcome the data sparsity.} Knowledge graph-based RSs can overcome the data sparsity and cold start problems faced by traditional RSs. The integration of rich domain knowledge and semantic information in the knowledge graph provides the RS with a more comprehensive understanding and modeling capability. This enables the system to perform better when dealing with new domains, long-tail items, and complex relationships.

\subsubsection{Methods}
Knowledge graph representation learning is a fundamental task in the field of artificial intelligence, with numerous applications ranging from information retrieval to RSs. One common challenge faced in this domain is the sparsity of data, which hinders the effectiveness of traditional approaches for completing the knowledge graph. To address this issue, researchers have explored various strategies, including embedding-based methods, path-based or rule-based methods, GNN-based methods, and pre-training-based methods.

\textbf{Embedding-based methods.} The embedding-based approach is a widely employed and potent technique for knowledge graph representation learning. It strives to discover low-dimensional embeddings that represent entities and relations within a semantic space, with the goal of capturing semantic associations between entities and the semantic interpretations of relations. The representation of each triple ($e_1$, $r$, $e_2$) in the semantic space is determined by a scoring function, which continuously adjusts the embeddings of entities and relations based on their co-occurrence relationships, learning properties like transitivity, symmetry, and invertibility between triplets.

TranSparse \cite{ji2016knowledge} and ComplEx \cite{trouillon2016complex} are both effective models tailored for dealing with sparse data. TranSparse improves upon the TransR method \cite{lin2015learning}. TranSparse goes a step beyond TransR by constraining the mapping matrix's expressive capacity to mitigate issues related to overfitting. Specifically, TranSparse introduces constraints in the mapping matrix to allow the model to better adapt to the diverse distribution of relations. Moreover, considering the imbalance in the distribution of entities, TranSparse takes into account that the occurrence frequency of the same entity in the head and tail positions may differ. Therefore, it learns distinct embedding representations for the same entity at different positions, further enhancing the model's ability to represent sparse data. To implement these constraints, TranSparse uses sparse matrices, effectively handling the sparse data problem in knowledge graphs. ComplEx, from another aspect, is a model based on complex-valued embeddings, representing both entities and relations as complex vectors. This complex representation offers more flexibility in capturing the semantic relationships between entities and relations, making it advantageous for handling sparse data. Similar to TranSparse, ComplEx also efficiently models sparse data in knowledge graphs through its simple yet effective approach.

\textbf{Path-based methods.} Path-based or rule-based methods make use of essential pathways connecting entity pairs ($e_1$, $e_2$) to represent the relationship $r$. For example, for the fact ($a$, $r_1$, $c$), there might exist a key path $a$-$r_2$-$b$-$r_3$-$c$ (corresponding to the facts ($a$, $r_2$, $b$) and ($b$, $r_3$, $c$)). Usually, there are multiple paths, and these paths (usually represented by the sequence of relations) collectively describe the relationship $r$. This class of methods provides some interpretability at the feature construction level, where the input consists of relationship predicate features composed of key paths, and the output is the confidence score. Moreover, there are some models based on multi-hop reasoning that can perform reasoning directly on the graph, providing complete logical paths and stronger interpretability. However, sparse data limits the richness of features in such approaches. The number of entity-to-entity paths obtained through sampling may significantly decrease, and there might be cases where no reachable paths can be found.

To address challenges, DacKGR uses an embedding model for multi-hop reasoning, providing extra paths through dynamic completion \cite{lv2020dynamic}. The model selects top-$k$ actions from the embedding model as candidate actions for reinforcement learning. Entity embeddings offer semantic references for path sampling and the model's confidence guides reinforcement learning. ComplEx-N3 also uses embedding-based multi-hop reasoning and dynamic path completion \cite{lacroix2018canonical}. Using complex embeddings, it captures semantic relationships to handle sparse data. Reinforcement learning in ComplEx-N3 selects top-$k$ actions from embeddings, leveraging confidence for path selection, enriching knowledge, and yielding accurate results.

\textbf{The combination of embedding and path methods.} Combining both embedding and path-based or rule-based methods is indeed possible and can be beneficial. TRE focuses on the fundamental fact of transitive relationships \cite{zhou2019completing}. Unlike typical embedding models, TRE only learns embeddings for relationships. One reason is that entity embeddings are greatly affected by sparsity, making it challenging to learn good entity embeddings. Moreover, transitive reasoning does not explicitly require entity embeddings. Relationship embeddings alone are sufficient. TRE models the learning of embeddings as learning transitive relationships, akin to a simplified version of rule learning, specifically learning two-hop rules. TRE introduces simple first-order logic by avoiding the complexity of learning entity embeddings. However, the model operates only on two-hop first-order logic, leading to some limitations. Nevertheless, this incremental step represents progress compared to dealing with rules of variable length and indirectly indicates that embedding-based learning can benefit tasks such as rule mining \cite{yin2022constraint}.

In addition, IterE incorporates axioms of relationships into the model learning process \cite{zhang2019iteratively}. It uses learned relationship axioms to augment the dataset, improving model performance on sparse data. Axioms here refer to rules for rules, encompassing various types of rules summarized into seven basic axioms. The process involves three steps: embedding learning, axiom induction, and axiom injection. Embedding learning involves training a complete embedding model using the original data. Axiom injection infers new knowledge using the current pool of axioms and incorporates it into the training set as new data.

\textbf{GNN-based method.} Graph neural networks (GNNs) have emerged as a potent category of neural networks designed to effectively process data with graph-like structures. They are particularly valuable in the context of knowledge graphs, which are represented as directed graphs with entities as nodes and relationships as edges. GNNs play a crucial role in knowledge graphs by utilizing feature vectors or embeddings associated with each entity and relationship. These feature vectors capture the semantic information of entities and relationships in a lower-dimensional space. By leveraging both the feature vectors and the graph's structure, GNNs can learn and propagate information throughout the graph. This propagation process enables GNNs to capture complex relational patterns and perform reasoning tasks effectively.

ExpressGNN enhances GNNs by introducing expressive substructures, such as shortest paths and neighborhood structures, into the framework \cite{zhang2020efficient}. Leveraging these subgraphs enriches the representation of nodes and relationships in the knowledge graph, enabling better capture of latent relationships and more effective information propagation. This improved expressiveness empowers ExpressGNN to handle sparse data and enhance knowledge graph reasoning. Similarly, pGAT addresses sparsity using attention mechanisms \cite{harsha2020probabilistic}. By incorporating graph attention at the path level between entities, pGAT weighs the importance of different paths in capturing relationship semantics. It focuses on informative paths while filtering out noise, thus improving the model's ability to reason over knowledge graphs in the presence of sparse data.

\textbf{Pre-training knowledge graph.} The use of pre-training methods is an effective approach to tackle the challenges posed by sparse data in knowledge graph completion tasks. Automatic KB completion for commonsense knowledge graphs leverages pre-trained models (PTMs) such as BERT \cite{malaviya2020commonsense}. The nodes in knowledge graphs often have general semantic meanings and frequently appear in natural language texts. By using PTMs, the model can directly incorporate this knowledge to aid in node encoding. The node representations that are more aligned with the semantics of natural language can be leveraged to provide prior knowledge. For instance, nodes with similar semantics in natural language can be connected with ``sim" edges to enrich the semantic information.

K-BERT is another pre-trained method specialized for knowledge graphs \cite{liu2020k}. During pre-training, K-BERT utilizes both masked language modeling tasks and knowledge graph completion tasks for joint learning. This allows K-BERT to harness the knowledge graph's information while learning natural language representations. The joint pre-training enables K-BERT to make full use of knowledge graph information, compensating for missing data in the knowledge graph and adapting better to sparse data situations. After pre-training, K-BERT can be applied to knowledge graph completion tasks. It is capable of inferring missing triple information based on known entities and relationships, thereby filling in the gaps in the knowledge graph. Due to the incorporation of knowledge graph information during pre-training, K-BERT can perform more accurate reasoning and completion, making it practical for dealing with sparse data scenarios.

\section{Challenges}  \label{sec:challenges}

\begin{figure}[ht]
    \centering
    \includegraphics[clip,scale=0.3]{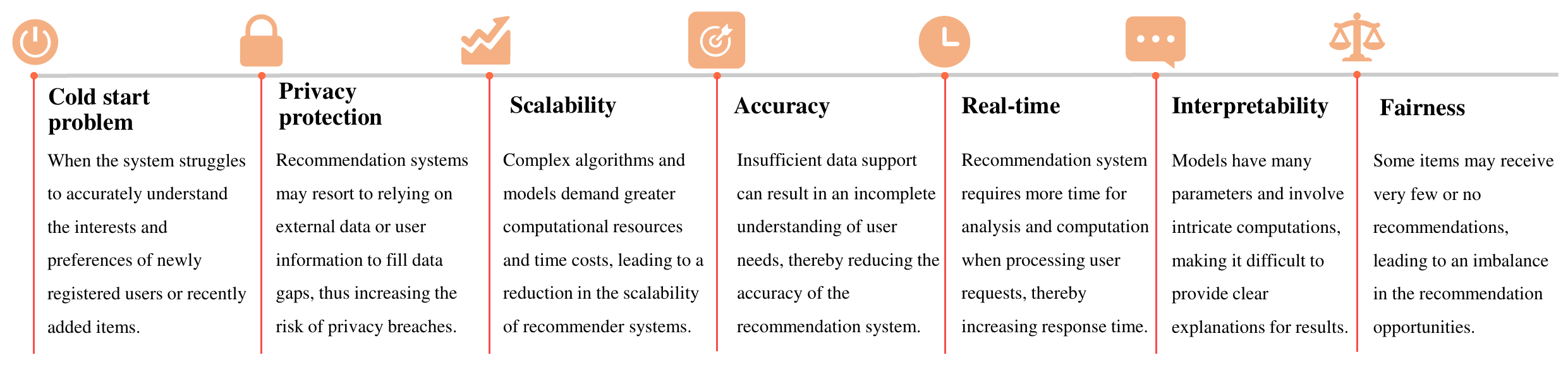}
    \caption{Challenges of recommendation system.}
    \label{fig:challenge}
\end{figure}

Data scarcity is a prevalent issue encountered by recommendation systems (RSs) in diverse real-world situations, which hinders the ability of the RS to obtain sufficient and diverse user behavior data and item information. In this section, we discuss the challenges posed by data scarcity in RSs. They are mainly divided into seven aspects, as illustrated in Fig. \ref{fig:challenge}.

\subsection{Cold start problem}

The cold start problem \cite{lam2008addressing}, as previously highlighted, plagues recommendation systems by hampering the system's ability to effectively cater to newly registered users or recently introduced items. While various strategies have been introduced to mitigate the cold start issue, recent efforts have specifically aimed at leveraging innovative solutions within recommendation systems. These methods include harnessing the potential of data augmentation, self-supervised learning, transfer learning, broad learning, and knowledge graphs. Each of these approaches offers distinct advantages in addressing the cold start conundrum. While effective, these strategies encounter challenges within the context of data scarcity and the cold start problem. Despite their utility, limitations persist. Data augmentation may struggle to accurately represent diverse new entities, affecting system generalization. Self-supervised learning might face difficulties extracting nuanced patterns from limited historical data. Transfer learning may encounter domain mismatch issues, and broad learning might struggle to manage diverse data sources effectively. Knowledge graphs may face hurdles in representing evolving entity connections. Addressing these challenges is pivotal for optimizing recommendation systems for new entities amidst data scarcity \cite{fan2022sequential, cao2023multi}.

\subsection{Privacy protection}

From a recommendation technology perspective, many recommendation techniques require the acquisition and analysis of sensitive user data, such as geographical location, preferences, and application data \cite{zhu2014mobile}, to provide more accurate recommendations. However, processing user privacy data may result in a reduced volume of available data, consequently impacting the effectiveness of recommendations. Particularly in cases of data scarcity, where sufficient data is lacking to generate precise personalized recommendations, RSs may resort to relying on external data or user information to fill data gaps, thus increasing the risk of privacy breaches. Under the backdrop of data scarcity, RSs primarily face two challenges. Firstly, they need to implement effective protection mechanisms to ensure the security of user data, preventing unauthorized access and misuse of personal information. Secondly, these systems need to find a balance between personalized recommendations and privacy protection, mitigating the impact of data privacy processing on recommendation performance. As a result, achieving accurate personalized recommendations through appropriate privacy protection mechanisms poses a challenging research problem. A Federated RS is an innovative approach in the field of RSs that addresses the challenges of data privacy and distribution. In traditional RSs, user data is typically aggregated and stored in a centralized manner, raising privacy concerns and potential security risks. Federated RS, on the other hand, adopts a decentralized architecture where data remains distributed across various local sources or entities while still enabling collaborative recommendation \cite{liu2022federated}.

\subsection{Scalability}

In RSs, scalability \cite{singh2020scalability} is crucial, as they need to process increasing amounts of data and computations to generate personalized recommendations, especially as the number of users and items grows. In the context of data scarcity, scalability implies that the RS faces challenges due to insufficient data when processing user behavior data and item information, thereby limiting its scalability. To address data scarcity, the RS may require more sophisticated algorithms and models to compensate for the lack of data and deliver accurate personalized recommendations. However, these complex algorithms and models demand greater computational resources and time costs, leading to an increase in the computational complexity of the RS and consequently reducing its scalability.

\subsection{Accuracy}

Usually, accuracy is all we need for recommendation systems. In practical applications, the accuracy of a RS is crucial for the user experience and business value \cite{jannach2021recommender}. A precise RS can enhance user satisfaction and trust in the recommended results. The accuracy of a RS refers to the degree of match between the recommended results and the user's true interests and needs. It is a significant performance metric that reflects the system's ability to predict user preferences and interests effectively. However, data scarcity can lead to a decrease in the prediction capability of the RS regarding user interests and preferences. Insufficient data support can result in an incomplete understanding of user needs, thereby reducing the accuracy of the RS.

\subsection{Real-time}

Real-time capability in a RS refers to its ability to swiftly generate and provide personalized recommendations after receiving a user's request. Specifically, it entails processing and responding to user queries within a short timeframe, typically ranging from milliseconds to seconds. Real-time \cite{chandramouli2011streamrec} performance is critical for RSs, particularly in scenarios involving real-time interactions and instant feedback. However, data scarcity presents challenges and impacts the real-time performance of RSs. Firstly, data scarcity increases computational complexity, which can hamper the real-time capability of the system. Secondly, the lack of sufficient user behavior data due to data scarcity may hinder the system from rapidly producing accurate recommendations. As a result, the RS requires more time for analysis and computation when processing user requests, thereby increasing response time. Lastly, in the context of data scarcity, system resources may become constrained, further intensifying the demand for system resources and potentially compromising real-time performance.

\subsection{Interpretability}

In RSs, the use of increasingly complex deep learning and machine learning models has often posed a challenge to interpretability \cite{vlachos2018addressing}. These models typically have many parameters and involve intricate computations, making it difficult to provide clear and straightforward explanations for their recommendation results. Data sparsity further exacerbates the challenge of interpretability for RSs. Firstly, data sparsity makes it challenging for RSs to offer detailed explanations for their recommendation results. Additionally, in the presence of data sparsity, RSs may need more complex algorithms and models to compensate for the effects of missing data. These intricate models may struggle to provide explanations to users about their recommendation process, thus reducing the overall interpretability of the RS. In the context of data scarcity, it becomes even more crucial to enhance interpretability to help users better comprehend the recommendation results and process, ultimately increasing user satisfaction and trust.

\subsection{Fairness}

In the digital economy, fairness \cite{wang2023survey} is an essential societal value, particularly in domains involving sensitive information such as job opportunities, financial services, and social resource allocation. In the presence of data scarcity, the fairness of RSs is more susceptible to impact, leading to various fairness challenges and issues. For instance, RSs may tend to prioritize recommending popular and widely-accepted items, as they often have more interaction behaviors and rating information. This may result in the system favoring mainstream groups and neglecting the interests of niche users, thus reducing fairness. Moreover, due to data scarcity, some items may receive very few or no recommendations, leading to an imbalance in the recommendation opportunities for different items. Some items may consistently lack exposure and recommendation opportunities, impacting opportunity fairness \cite{sonboli2022multisided}.

\section{Future Directions} \label{sec:direction}

Here, we present an overview and future directions for studying data scarcity in recommendation systems (RSs). There are certain issues with current experimental research on user habits, and it is important to explore future research directions.

\subsection{Multi-modal recommendation}

Multi-modal recommendation \cite{wei2019mmgcn, liu2019joint} is a technique that leverages multiple types of data to enhance the accuracy and coverage of RSs. Integrating various data types, such as user text, images, audio, etc., can enrich the descriptions of users and items. Through the utilization of this multimodal data, RSs can gain a more comprehensive understanding of users' interests and preferences, leading to more precise recommendations. In the context of data scarcity, multimodal recommendation offers a crucial solution for RSs. To overcome the challenges posed by data scarcity, further research, and innovation are needed in multimodal recommendation techniques. The development of more efficient and accurate fusion models and algorithms becomes essential to effectively supporting and optimizing RSs.

\subsection{Metaverse recommendation}

In the Metaverse era \cite{sun2022metaverse,sun2022big,chen2022metaverse}, there are many potential future directions for RSs. In the Metaverse era, there are many potential future directions for RSs. RSs play a vital role in safeguarding user experiences within a human-centric Metaverse \cite{chen2023open,yang2023human}. RSs can play various roles in enhancing user experiences and interactions in the Metaverse. These roles include providing highly personalized experiences, recommending virtual items and assets, facilitating social interactions, guiding virtual tourism and exploration, suggesting virtual education and training, offering virtual entertainment and media, promoting virtual advertising, and ensuring safety and privacy. In this diverse and immersive virtual environment, recommendation systems aim to improve user engagement and contribute to the growth of the Metaverse.

\subsection{Context-aware recommendation}

The future directions of recommendation systems are poised for diverse innovations, particularly in the realm of context-aware recommendation \cite{adomavicius2010context}, presenting some exciting prospects. This field focuses on a deeper understanding of user context, and environment, and needs to offer more personalized and contextually relevant recommendations. Future trends in context-aware recommendation might explore perceptual intelligence and integration with social networks. Integrating natural language processing, computer vision, and sentiment analysis to enhance the system's understanding of user emotions, attitudes, and context. This implies a more accurate capture of user sentiments, interests, and preferences for delivering emotionally intelligent recommendations. In social networks, deeper integration of social network data to harness user interactions and behaviors on social platforms. This integration can comprehensively understand user backgrounds and requirements, leading to more targeted recommendation services.

\subsection{Recommendation based on Meta-learning and adaptive algorithms}

A future direction for RSs is a recommendation based on meta-learning and adaptive algorithms. This direction aims to enhance the personalization and adaptability of RSs, enabling them to better meet the needs and interests of users.

Meta-learning \cite{vilalta2002perspective} is a machine-learning approach that improves the performance of learning algorithms by learning how to learn. In the context of RSs, meta-learning can be used to learn how to select and adjust recommendation algorithms and models, for providing personalized recommendations based on the characteristics of different users and contexts. Through meta-learning, RSs can automatically learn and adapt to the preferences and behavior patterns of individual users, thereby delivering more accurate and personalized recommendations \cite{cui2016recommendation}.

Adaptive algorithms refer to algorithms that can automatically adjust based on the environment and user feedback. In RSs, adaptive algorithms can dynamically adjust and optimize based on user feedback and behavior \cite{cheng2021adaptive}. This includes real-time updates based on user clicks, purchases, ratings, and other actions, as well as considering factors such as time, location, and context for dynamic adjustments. Through adaptive algorithms, RSs can better adapt to user changes and evolving interests, providing more precise and timely recommendations. RSs based on meta-learning and adaptive algorithms offer the following advantages:

\begin{itemize}
    \item \textbf{Personalization.} Such RSs can provide customized recommendations based on users' personalized needs and interests, offering more personalized and accurate recommendations.

    \item \textbf{Adaptability.} RSs can dynamically adjust based on changes in different users and contexts, maintaining the timeliness and accuracy of recommendations.

    \item \textbf{Real-time responsiveness.} Adaptive algorithms enable real-time updates of recommendations, adjusting based on users' real-time feedback and behavior, providing timely and targeted recommendations.

    \item \textbf{Flexibility.} Meta-learning allows RSs to learn and select the most suitable recommendation algorithms and models, flexibly adjusting based on the characteristics of different users and contexts.
\end{itemize}

\subsection{Recommendation based on UGC and community engagement}

A future direction for RSs is based on user-generated content (UGC) \cite{krumm2008user} and community engagement \cite{head2007community}. This strategy aims to improve the recommendation process by utilizing user-generated content and their participation in online communities. UGC refers to any type of content that users rather than the platform itself create, such as reviews, ratings, comments, and social media posts. By incorporating UGC into the RS, valuable insights can be gained from the collective wisdom of the community, leading to more accurate and relevant recommendations. The benefits of RSs based on UGC and community engagement include but not limited to:

\begin{itemize}
    \item \textbf{Enhanced relevance.} Recommendations can be more aligned with the preferences and interests of users due to the collective knowledge and experiences shared within the community.

    \item \textbf{Discovery of niche content.} Users can discover unique and specialized content that may not be easily found through traditional recommendation approaches.

    \item \textbf{Increased user engagement.} By actively engaging users in the recommendation process and promoting community participation, users are more likely to actively participate and contribute to the platform.

    \item \textbf{Trust and credibility.} Recommendations based on UGC and community engagement can be perceived as more trustworthy and credible, as they are influenced by real users' experiences and opinions.
\end{itemize}

However, challenges exist in implementing RSs based on UGC and community engagement. These challenges include managing the quality and reliability of UGC, addressing biases and manipulation, and ensuring privacy and data protection.

\subsection{Trust-based recommendation systems}

Trust-based RSs \cite{walter2008model, andersen2008trust} are designed to enhance the trustworthiness and credibility of recommendations by incorporating trust-related factors into the recommendation process. These systems aim to tackle the challenge of information overload and help users make informed decisions by considering trustworthiness as a crucial aspect.

In trust-based RSs, trust is typically defined as the subjective belief or confidence that a user has in another user, an item, or the RS itself. Trust can be established through various mechanisms, such as explicit ratings and reviews, social connections, past interactions, and reputation systems. There are some key components and characteristics of trust-based RSs, as follows:

\begin{itemize}
    \item \textbf{Trust modeling.} Trust-based RSs utilize mathematical models or algorithms to quantify and represent trust relationships between users, items, or the RS. These models can take into account various trust factors, such as reliability, expertise, similarity, and social influence.

    \item \textbf{Trust propagation.} Trust information can be propagated or spread through a network of users or items. As an illustration, if User A places their trust in User B, and User B, in turn, expresses trust in User C, this trust in User C can be extended through the trust network that connects User A and User B. This propagation helps in assessing the trustworthiness of unknown or unfamiliar items or users.

    \item \textbf{Trust-based filtering.} Trust-based RSs use trust information to filter and rank the recommendations. Recommendations from highly trusted sources or users are given more weight or priority. This filtering process aims to provide users with recommendations that are more likely to be reliable and align with their preferences.

    \item \textbf{Trust-aware aggregation.} Trust-based RSs consider trust information when aggregating or combining multiple recommendations. The trustworthiness of individual recommendations can influence the overall aggregated recommendation. This approach helps mitigate the impact of potentially biased or malicious recommendations.

    \item \textbf{Trust visualization.} Trust-based RSs may provide visual representations or indicators of trust to users. These visualizations help users understand the trustworthiness of recommendations and make informed decisions. For instance, a recommendation might come with a trust score or a visual representation of the trust level.
\end{itemize}

The benefits of trust-based RSs include enhanced credibility, improved decision-making, and personalization. By considering trust-related factors, these systems can provide recommendations that are perceived as more trustworthy and reliable by users. Trust-based recommendations help users navigate through the abundance of information and make informed choices based on trusted sources or users. Trust models can be personalized to reflect individual users' trust preferences and relationships, leading to more personalized and relevant recommendations.

\section{Conclusions} \label{sec:conclusions}

In this article, we delve into the intricate issue of data scarcity within recommendation systems (RSs). Through meticulous analysis, we have identified the intricate challenges stemming from limited data availability and extensively explored diverse methodologies to alleviate its impact. Our article underscores the significance of cultivating robust techniques capable of adeptly navigating data scarcity, encompassing strategies such as data augmentation, self-supervised learning, transfer learning, broad learning, and harnessing knowledge graphs. By implementing these solutions, researchers and practitioners can substantially elevate the performance and precision of RSs, even within constrained data environments. Our investigation seamlessly intertwines with the concept of knowledge transfer. By integrating knowledge from external sources, RSs are empowered to surmount data scarcity hurdles effectively. Furthermore, our research extends beyond solutions, shedding light on the pressing challenges and promising future trajectories regarding data scarcity in RSs. We aim to furnish a comprehensive compendium for tackling data scarcity quandaries in the realm of RS development. In doing so, we provide a wellspring of insights and guidance to propel the progress of both researchers and developers dedicated to advancing the landscape of RSs.

\section*{Acknowledgment}

This research was supported in part by the National Natural Science Foundation of China (Nos. 62002136 and 62272196), Natural Science Foundation of Guangdong Province (No. 2022A1515011861), the Young Scholar Program of Pazhou Lab (No. PZL2021KF0023), Engineering Research Center of Trustworthy AI, Ministry of Education (Jinan University), and Guangdong Key Laboratory of Data Security and Privacy Preserving.

\bibliographystyle{ACM-Reference-Format}
\bibliography{paper.bib}
\end{document}